\documentclass[aps,prb,superscriptaddress,twocolumn,showpacs]{revtex4}
\usepackage{graphicx}

\newcommand{\be}{\begin{equation}}
\newcommand{\ee}{\end{equation}}
\newcommand{\bea}{\begin{eqnarray}}
\newcommand{\eea}{\end{eqnarray}} 
\newcommand{\edip}{E_{\hbox{\tiny dip}}}  
  
\newcommand{\Weff}{W_{\hbox{\tiny eff}}}  
\newcommand{\leff}{\lambda{\hbox{\tiny eff}}}  
\newcommand{\eani}{E_{\hbox{\tiny ani}}}  
\newcommand{\hdip}{H_{\hbox{\tiny dip}}}  
\newcommand{\dmin}{D_{\hbox{\tiny min}}}  
\newcommand{\su}{\vec s_1}
\newcommand{\sd}{\vec s_2}
\newcommand{\Su}{\vec{\cal S}_1}
\newcommand{\Sd}{\vec{\cal S}_2}
\newcommand{\fra}[2]{\hbox{${#1\over #2}$}} 

\begin{document}

\title{Dipolar interaction between two-dimensional magnetic particles}

\author{Paolo Politi}
\altaffiliation{Corresponding author.\\Electronic address:
{\tt politi@ifac.cnr.it}}
\affiliation{Istituto di Fisica Applicata ``Nello Carrara",
Consiglio Nazionale delle Ricerche, Via Panciatichi 56/30,
I-50127 Firenze, Italy}
\affiliation{Istituto Nazionale per la Fisica della Materia,
UdR Firenze, Via G. Sansone 1, I-50019 Sesto Fiorentino, Italy}

\author{Maria Gloria Pini}
\email{mgpini@ifac.cnr.it}
\affiliation{Istituto di Fisica Applicata ``Nello Carrara",
Consiglio Nazionale delle Ricerche, Via Panciatichi 56/30,
I-50127 Firenze, Italy}

\date{\today}

\begin{abstract}
We determine the effective dipolar interaction between single domain 
two-dimensional ferromagnetic particles (islands or dots), 
taking into account their
finite size. The first correction term decays as $1/D^5$, where $D$ is
the distance between particles.
If the particles are arranged in a regular two-dimensional array and 
are magnetized in plane, we show that the correction term reinforces  
the antiferromagnetic character of the ground state in a square lattice,
and the ferromagnetic one in a triangular lattice.
We also determine the dipolar spin-wave spectrum and 
evaluate how the Curie temperature
of an ensemble of magnetic particles scales with the parameters
defining the particle array: height and size of each particle, and 
interparticle distance. 
Our results show that dipolar coupling between particles might
induce ferromagnetic long range order at experimentally relevant 
temperatures. However, depending on the size of the particles,
such a collective phenomenon may be disguised by superparamagnetism.
\end{abstract}


\maketitle

\section{Introduction}

In this paper we are interested in two-dimensional magnetic particles
interacting through the long range dipolar forces.
These particles have a two-dimensional character in two respects:
first, because they are platelet shaped, that is to say their 
thickness $t$ is much smaller than their linear size $L$;
second, because they are arranged on a two dimensional substrate.

They can be obtained, {\it e.g.}, growing by Molecular Beam Epitaxy (MBE) 
a magnetic element on a high symmetry substrate.
In this case, growth is driven by surface diffusion, nucleation 
and aggregation:~\cite{growth_mbe} in the submonolayer regime,
the magnetic overlayer is made up of an ensemble of atomically
thick islands that generally are not uniform in size neither
arranged in a regular array.
The distribution of islands may be regular if nucleation
(and therefore island formation) takes place 
on a reconstructed surface,~\cite{ricostruzione} or on a network of
dislocations.~\cite{dislocazioni} Alternatively, 
particles can be produced via lithographic techniques:\cite{litografia}
in this case, they are much bigger in size and their
distribution is generally uniform.

In the following, small particles obtained by MBE growth in the
submonolayer regime will also be called {\em islands} and 
large particles obtained by lithographic techniques will also be
called {\em dots}. {\em Particle} is a generic term for both cases.

Each particle is made up of a large number ${\cal N}$ of spins 
which interact 
ferromagnetically through the strong intra-particle exchange interaction. 
In an island, ${\cal N}\approx 10^2-10^4$, while in a dot
${\cal N}$ may be several orders of magnitude greater.

Sufficiently small particles are expected to be in a single domain
state, even if their actual magnetic state may depend on several
factors: the shape of the particle, the strength of the anisotropies,
the single crystal or polycrystalline character of the particle,
and so on. In this paper we are assuming that particles are
in a single domain state and have a crystalline structure.
Within these hypotheses,
the magnetic state of an isolated particle is fixed, first of all,
by the balancing between dipolar interaction (which has an
easy-plane effect) and possible anisotropies
favouring the direction perpendicular to the plane ($z$ direction,
in the following). In the absence of quartic and higher order
anisotropies a canted configuration is impossible and the resulting 
effect may be easy-axis or easy-plane only. 

The effect of dipolar interaction between spins belonging to the same 
(ultrathin) particle has been studied in a previous paper.~\cite{stripes}
We showed that in-plane shape anisotropy is weak:
this means that if the magnetization of
the particle is within the film plane, its orientation is expected to be
settled by the symmetry of the underlying lattice 
(through magnetocrystalline anisotropies) rather than by the shape of the 
particle (through the intra-particle dipolar coupling). 
Such a feature was indeed experimentally observed 
in MBE-grown Co on Cu(100) ultrathin particles.\cite{Science}

In this paper we aim to study the interparticle dipolar
interaction and to address the following questions:

{\it i)} Assuming the particle to be in a single domain state, 
how is the dipolar interaction between particles
modified by their finite size?

{\it ii)} Assuming the particles to be arranged in a regular array,
is the dipolar ferromagnetic (FM) state a stable configuration?

{\it iii)} If the ground state is ferromagnetic, is it stable at finite
temperature and what is the value of the Curie temperature?

The previous questions are addressed in Sections~\ref{sec:int_eff},
\ref{sec:spin-wave} and \ref{sec:curie_temp}, respectively;  
in Section~\ref{sec:concl} the conclusions are drawn.

\section{Effective dipolar interaction between particles}
\label{sec:int_eff}

In the following we are going to consider two particles 
$I_1,I_2$ of any shape (see Fig.~1), with linear sizes 
$L_1,L_2$ and thicknesses $t_1,t_2$.
Each particle is a discrete
collection of spins and it is supposed to be
in a single domain state:
each spin of the two particles is indicated by 
$\su$ and $\sd$, respectively.
The effective dipolar interaction between the two particles is
evaluated by taking the interaction between a spin $\su$ located
in $\vec R_1$, a spin $\sd$ located in $\vec R_2$ and
summing up on all them: 

\be
\edip = {1\over 2}\Omega {t_1 t_2\over c_0^2} 
\sum_{\vec R_1} \sum_{\vec R_2}
\left[
{ \su \cdot \sd\over R_{12}^3 } 
-3 { (\su\cdot\vec R_{12})
(\sd\cdot\vec R_{12})\over R_{12}^5 } \right]
\label{edip}
\ee
\noindent where $\Omega=g^2\mu_B^2$ ($g$ is the gyromagnetic factor
and $\mu_B$ the Bohr magneton) and $c_0$ is the interplane distance in the
$z$ direction.
In the previous expression, we have supposed that the thickness 
of each particle is much smaller than its linear size, $t\ll L$.
In this hypothesis, $\edip$ is just linear in the
numbers of atomic planes, $t_1/c_0$ and $t_2/c_0$,
and $\vec R_{12}$ are two dimensional vectors.

\begin{figure}
\includegraphics[width=8cm]{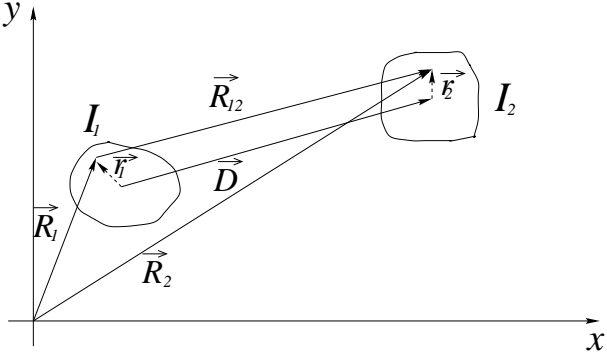}
\caption{$I_1$ and $I_2$ are two generic interacting particles.
Each spin of the first particle, located in $\vec R_1$, interacts with
each spin of the second particle, located in $\vec R_2$.
Using reference frames centered in the centers of mass of the particles,
the spatial positions of the two spins are $\vec r_1,\vec r_2$ 
so that the distance between the two spins can be expressed as 
$\vec R_{12}=\vec R_2 -\vec R_1 = \vec D +\vec r_2-\vec r_1
\equiv \vec D +\vec r$. }
\end{figure}

As explained in Appendix~\ref{app1}, the two quantities appearing in
square brackets in Eq.~(\ref{edip})
can be expanded in the ratio $r/D$, where $\vec D$
is the distance between 
the centers of mass of the particles and $\vec r=\vec r_2-\vec r_1$
(see Fig.~1).
The effective dipolar interaction between two particles at distance $\vec D$
takes the approximate form:
\be 
\edip \approx \edip^{(0)} + \edip^{(2)}
\ee
where $\edip^{(0)}$ is the zero-order coupling and $\edip^{(2)}$ takes
into account the finite size of the particles:
\be
\edip^{(0)} ={1\over 2} \Omega \left[ { \Su\cdot\Sd \over D^3} -3
{ (\Su\cdot\vec D)(\Sd\cdot\vec D) \over D^5} \right]\label{e0} 
\ee
\bea
\edip^{(2)} &=& \Omega {9 {\cal I}_{12}\over 4} 
{ \Su^\perp\cdot\Sd^\perp \over D^5} \nonumber\\
&& + \Omega {3 {\cal I}_{12}\over 4} \left[
{ \Su^\parallel\cdot\Sd^\parallel \over D^5} 
-5 { (\Su^\parallel\cdot\vec D)(\Sd^\parallel\cdot\vec D) \over D^7} \right] 
\label{e2}
\eea

Each particle behaves as a single spin $\vec{\cal S}={\cal N}\vec s$,
where ${\cal N}$ is the total number of spins in the particle. If $L$ is the 
linear dimension of a particle and $t$ its thickness, 
denoting by $a_0$ the in-plane atomic distance,
one has ${\cal N} = \gamma (L/a_0)^2 (t/c_0)$, where $\gamma$ is a geometric
factor, depending on the shape of the particle and the lattice structure.
The correction terms included in $\edip^{(2)}$ decay with
distance as $1/D^5$, whilst the usual dipolar interaction
decays as $1/D^3$. More precisely, $\edip^{(2)}$ 
is a factor ${\cal I}_{12}/D^2$ smaller than $\edip^{(0)}$,
where ${\cal I}_{12}=\fra{1}{2}({\cal I}_1+{\cal I}_2)$ is the semisum
of the ``moments of inertia" ${\cal I}_i$ of the two particles (see 
Appendix~\ref{app1}). 

In the continuum approximation, we have the following expressions.
For a square particle of side $L$, ${\cal I}=L^2/6$; 
for a circular particle of radius $\rho$, ${\cal I}=\rho^2/2$; 
for a triangular (equilateral) particle of side $L$, ${\cal I}=L^2/12$.

Finally, we would like to remark that the isotropic coupling term,
proportional to $(\Su\cdot\Sd)$, has different correction terms
according to the orientation of the spins: see Eq.~(\ref{e2}).
In other words, such a term is no more isotropic once the finite size
of the particle is taken into account.

In Fig.~2 we compare the exact dipolar coupling $\edip$ (symbols)
with the zero order approximation $\edip^{(0)}$ (dashed lines)
and with the second order approximation $(\edip^{(0)}+\edip^{(2)}$)
(full lines), in two cases: i)~the particles are magnetized 
perpendicularly to the plane, 
along the $\hat z$ axis (positive coupling energies), and
ii)~the particles are magnetized in plane
along the $\hat x$ axis (negative coupling
energies). 
We can see that the second order approximation is fairly good
except at very small distances: the smallest allowed distance between
centers (without superposing the particles)
is $\dmin = \sqrt{2}(L+a_0) = 29.7a_0$, where $L=20a_0$ 
is the side of the square particle. The smallest value of $D$ plotted in 
Fig.~2 is $D=30a_0$.

\begin{figure}
\includegraphics[angle=-90,width=8cm]{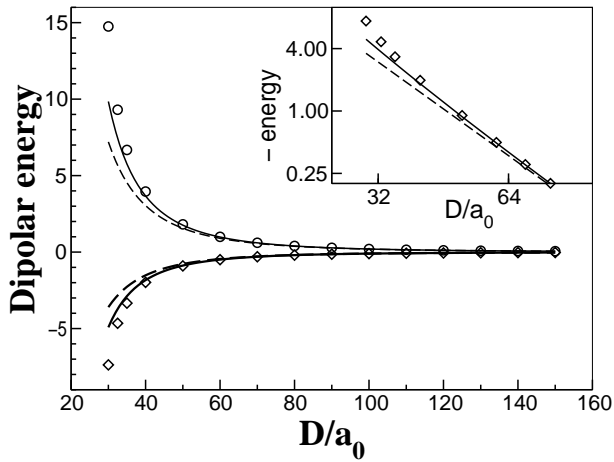}
\caption{Dipolar energy between two square particles,
one monolayer thick,  of side $L=20 a_0$
on a square lattice (21 spins per side). The centers of the two
particles have the coordinates (0,0) and $(D/\sqrt{2},D/\sqrt{2})$.
The sides of the squares are parallel to the axis $\hat x,\hat y$.
Positive and negative energies refer to spins parallel to $\hat z$ and
$\hat x$, respectively.
The exact calculation $\edip$ (symbols) is compared with
$\edip^{(0)}$ (dashed lines) and $(\edip^{(0)}+\edip^{(2)}$)
(full lines). Inset: the case of spins parallel to $\hat x$,
in a log-log scale. 
}
\end{figure}

In order to understand which configurations are energetically
favoured, let us start by considering just a couple of spins
$\vec S_1,\vec S_2$, located in plane along the $\hat x$ axis.
The dipolar coupling can be generally written as 
\bea
E_{12} &\approx& \phantom{+}E_{12}^{(0)} + E_{12}^{(2)} \nonumber\\
&\approx& \phantom{+}\tilde\Omega_0 (\vec S_1\cdot\vec S_2 -3 S_1^x S_2^x )\nonumber\\
&& +\tilde\Omega_2 (2S_1^z S_2^z + \vec S_1\cdot\vec S_2 -5 S_1^x S_2^x )
\nonumber
\eea  
where the explicit expressions of $\tilde\Omega_0$ 
and $\tilde\Omega_2$,
depending on the distance between spins and on the size of the particles,
are irrelevant. We observe that 
$\vec S_1,\vec S_2$ may be either veritable spins $\vec s_1, \vec s_2$
or they may represent the effective spins of two particles 
$\vec {\cal S}_1,\vec {\cal S}_2$:  in the former case one has 
$\tilde\Omega_2=0$, while in the latter $\tilde\Omega_2$ is the 
correction due to the finite sizes of the particles.

Both $E_{12}^{(0)}$ and $E_{12}^{(2)}$ are the sum
of competitive interactions. For the sake of definiteness, let
us consider $E_{12}^{(2)}$. The sum $2S_1^z S_2^z + \vec S_1\cdot\vec S_2$
favours an antiferromagnetic (AFM) alignment of the spins in the $\hat z$
direction, perpendicular to their joining vector. 
The term $-5 S_1^x S_2^x$ favours a ferromagnetic (FM) alignment
along the $\hat x$ axis. The energy of the former configuration is
$-3\tilde\Omega_2$ and the energy of the latter one is 
$-4\tilde\Omega_2$.
As proved in Appendix~\ref{app2}, where a more detailed discussion is given,
the latter configuration is the ground state indeed.

In conclusion, two spins interacting through the dipolar coupling
minimize their energy by ordering ferromagnetically along the
joining line. 

We are now going to discuss the more complex case 
of a two dimensional lattice of spins.
It is well known~\cite{Maleev} 
that in the presence of a direct exchange interaction,
the system is ferromagnetic and magnetized {\em in} the plane,
because of an easy-plane effect of $\edip$. If such an exchange
interaction is absent, the easy-plane effect survives, but the
actual configuration in the plane strongly depends on the lattice
structure.\cite{Roz91} It is useful to explain the origin of 
such a dependence on the spin arrangement.

We have seen that spins would like to 
point along the line joining them: in a two dimensional
lattice it is impossible, of course, to fulfill this requirement for
all couples of spins. It is possible, however, for a chain of spins:
so, we can start by addressing the nature of the coupling between 
chains.\cite{Malo,Frae99}
Let us consider a ferromagnetic chain of spins along the 
$\hat y$ axis of the plane and evaluate the dipolar field $\vec\hdip$
generated at a point at distance $d$. Spins are oriented in the
$+\hat y$ direction and $\vec\hdip = \hdip\vec y$. 
In the continuum approximation, 
\bea
\hdip = &-&{1\over 2}\Omega{\cal S} \int_{-\infty}^{+\infty} dy
\left({1\over r^3} -3{y^2\over r^5}\right) \nonumber\\
&-&\fra{3}{4}\Omega{\cal S} {\cal I} \int_{-\infty}^{+\infty} dy
\left({1\over r^5} -5{y^2\over r^7}\right) \nonumber
\eea
where $r=\sqrt{y^2+d^2}$.

Both integrals have the form $\int dy (\fra{1}{r^{3+n}}
-(3+n)\fra{y^2}{r^{5+n}})$.
It is sufficient to integrate by parts the term $1/r^{3+n}$
to prove that the integral vanishes
for any $n$ and for any $d$.
Therefore, in the continuum approximation $\hdip=0$.

An exact calculation on a discrete lattice gives a finite value for
$\hdip$, but its sign depends on the actual lattice structure.
Therefore, the reason for the sensitivity of the ground state on the
spin arrangement is clear. Dipolar interaction favours 
the formation of spin chains magnetized ferromagnetically along the chain:
these chains are very weakly coupled, in the continuum approximation
being even uncoupled. The sign of the coupling and therefore
the nature of the ground state do depend on the lattice
structure.

In the next Section we are going to analyze the nature of the ground state
and to study the spin-wave spectrum with respect to a ferromagnetic alignment.

\section{Ground states and spin-wave spectra in two-dimensional lattices}
\label{sec:spin-wave}

\begin{figure}
\includegraphics[bbllx=86pt,bblly=111pt,bburx=560pt,bbury=720pt,
 width=8cm]{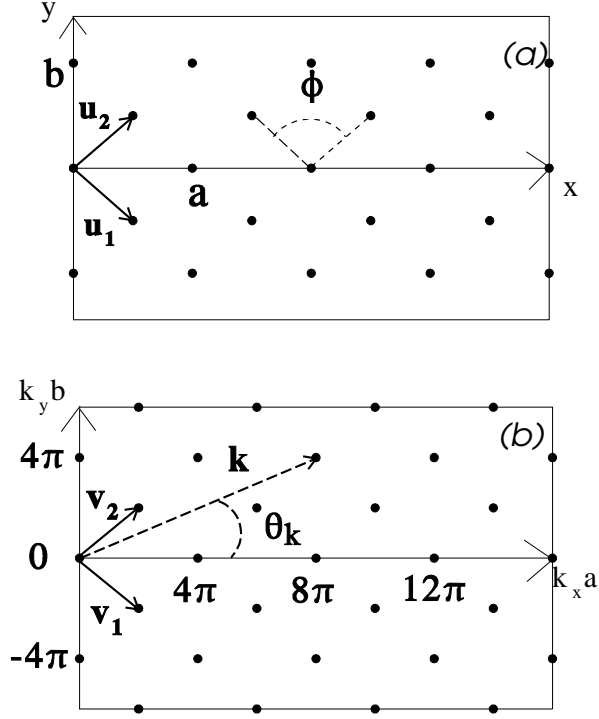}
\caption{(a) Orthorombic lattice with coordinate axes directed along
the rhombus diagonals, $a$ and $b$; $\phi$ is the rhombic angle.
(b) Reciprocal lattice of the orthorombic lattice.
The angle $\theta_k$ denotes the orientation of the
in-plane wavevector ${\vec k}$ with respect to the ${\hat x}$ axis.
}
\end{figure}

In the following we consider an orthorombic lattice
(see Fig.~3a) with primitive vectors 
\be
{\vec u}_1=\fra{1}{2}(a,-b),~
{\vec u}_2=\fra{1}{2}(a,b)
\ee
\noindent where $a=2 D_0 \sin (\phi/ 2)$, 
$b=2 D_0 \cos (\phi/ 2)$, and
$D_0$ is the rhombus side. 
For $\phi=\fra{\pi}{3}$ and $\phi=\fra{\pi}{2}$,
the orthorombic lattice reduces to a triangular and 
to a square lattice, respectively.
The angle $\phi$ can be supposed smaller than $\fra{\pi}{2}$, because if
$\phi>\fra{\pi}{2}$ there is just an interchange between $a$ and $b$
(see Fig.~3a). 

It is known~\cite{Roz91,Frae99} that the ground state is
ferromagnetic for $\phi < \phi_c\approx 80^\circ$ and antiferromagnetic
for larger values of the angle $\phi$.
It is noteworthy that we find $\phi_c$ to be
almost unaffected when $\edip^{(2)}$ 
is added to $\edip^{(0)}$: as a first approximation, 
the finite size of the particle has therefore
negligible effects on the ground state.
We can say as well that the finite size of the particles ({\it i.e.}
the term $\edip^{(2)})$ reinforces the FM character of the ground state
for a triangular lattice, $\phi={{\pi}\over 3}$,  and the AFM one
for a square lattice, $\phi={{\pi}\over 2}$.
This statement will be substantiated by the analysis of the spin-wave
spectra at the end of this Section. 

Let us now calculate the frequency of spin-wave excitations with respect 
to a ferromagnetic configuration with magnetization directed in plane.
We remind that, depending on the value of the rhombic angle
$\phi$, the ferromagnetic ground state is different:\cite{Roz91}
for $\phi\le \pi/3$, the dipoles are oriented along the 
short diagonal of the rhombus ({\it i.e.} along the $\hat x$ axis 
in Fig.~3a) while for $\pi/3 \le \phi \le \phi_c$, 
along the long diagonal ($\hat y$ axis). 

The effective dipolar interaction between particles of spins 
whose centers of mass are located on the sites $\vec D_{i}$ of 
an orthorombic lattice is 
${\cal H}_{\hbox{\tiny dip}}=
{\cal H}_{\hbox{\tiny dip}}^{(0)}+
{\cal H}_{\hbox{\tiny dip}}^{(2)}$; it is
obtained from Eqs.~(3,4) summing over 
all sites 
\bea
{\cal H}_{\hbox{\tiny dip}}^{(0)} &=& {1\over 2}
\Omega \sum_{\vec D_i} \sum_{\vec D_j}
\left[ { {\vec {\cal S}_i} \cdot {\vec {\cal S}_j} \over D_{ij}^3} -3
{ ({\vec {\cal S}_i}\cdot {\vec D_{ij}})
({\vec {\cal S}_j}  \cdot{\vec D_{ij}}) \over D_{ij}^5} 
\right]\label{h0}\nonumber  \\ 
{\cal H}_{\hbox {\tiny dip}}^{(2)} &=& 
{1\over 2}
\Omega \sum_{\vec D_i} \sum_{\vec D_j}
\Bigg\{
3 {\cal I}_{ij} \left[
{ {\vec {\cal S}_i}^\perp\cdot{\vec {\cal S}_j}^\perp \over D_{ij}^5} 
\right] \nonumber\\
&& + {3 {\cal I}_{ij}\over 2} \left[
{ {\vec {\cal S}_i}\cdot{\vec {\cal S}_j} \over D_{ij}^5} 
-5 { ({\vec {\cal S}_i}\cdot{\vec D_{ij}})
({\vec {\cal S}_j}\cdot{\vec D_{ij}}) \over D_{ij}^7} \right] 
\Bigg\} 
\label{h2}
\eea
\noindent where ${\vec D_{ij}}={\vec D_j}-{\vec D_i}$ and 
${\cal I}_{ij}={1\over 2}({\cal I}_{i}+{\cal I}_{j})$
is the semisum of the ``moments of inertia" of two particles whose
centers of mass are located in ${\vec D_{i}}$ and 
${\vec D_j}$, respectively. In the following we are assuming
to have an array of identical particles, so that ${\cal I}_i \equiv
{\cal I}$.

Taking the $\hat y$ direction as quantization axis, we perform 
the Holstein-Primakoff transformation from spin to boson operators
\be
{\cal S}_j^x=i\sqrt{{{\cal S}\over 2}}(a_j^{\dagger}-a_j^{}),~
{\cal S}_j^y={\cal S}-a_j^{\dagger}a_j^{},~
{\cal S}_j^z=\sqrt{{{\cal S}\over 2}}(a_j^{}+a_j^{\dagger})
\nonumber
\ee
\noindent Next, exploiting the translational invariance 
in the film plane, we introduce the Fourier transform
\be
a_j=\sqrt{{1\over N_{\Vert}}}\sum_{\vec k}
a_{\vec k} 
{\rm e}^{-i{\vec k}\cdot {\vec D}_{ij}},~
a_j^{\dagger}=\sqrt{{1\over N_{\Vert}}}\sum_{\vec k}
a^{\dagger}_{\vec k}
{\rm e}^{i{\vec k}\cdot {\vec D}_{ij}},~
\nonumber
\ee
\noindent where 
${\vec k}=(k_x,k_y)$ is the two-dimensional 
in-plane wavevector ranging over the first 
Brillouin zone, generated by the primitive vectors (see Fig.~3b)
\be
{\vec v}_1=2\pi (\fra{1}{a}, -\fra{1}{b}),~
{\vec v}_2=2\pi (\fra{1}{a}, \fra{1}{b})
\ee
\noindent and $N_{\Vert}$ is the total number of 
spins in the two-dimensional lattice.
The spin-wave Hamiltonian takes the form
\be
\label{spinwave}
{\cal H}_{\hbox{\tiny dip}}=\sum_{\vec k}
\left\lbrack A_{\vec k}
a^{\dagger}_{\vec k} a_{\vec k}
+ {1\over 2} 
B_{\vec k}
(
a_{\vec k} a_{-{\vec k}} +
a^{\dagger}_{\vec k} a^{\dagger}_{-{\vec k}} 
)
\right\rbrack
\ee
The coefficients 
$A_{\vec k}= A^{(0)}_{\vec k} + A^{(2)}_{\vec k}$ 
and
$B_{\vec k}= B^{(0)}_{\vec k} + B^{(2)}_{\vec k}$
can be expressed (see Appendix~C) through the dipolar sums
($\alpha,\beta=x,y,z$)
\be
{\cal D}^{(n)}_{\alpha \beta}({\vec k})=\sum_{{\vec D}_j} 
{1\over {D_{ij}^{3+n}}} \left[ 1- (3+n) 
{ {D_{ij}^{\alpha}D_{ij}^{\beta} }\over 
{D_{ij}^2}}
\right] 
{\rm e}^{i {\vec k} \cdot {\vec D}_{ij}}
\ee
where for $n=0$ and $n=2$ one has respectively the zero and 
second order expressions 
in the ratio $L/D_0$ between the linear dimension of the particle
and the interparticle distance.
The spin-wave energy is 
\be
\epsilon_{\vec k}=\left[
\Big( A_{\vec k}- B_{\vec k}\Big)
\Big( A_{\vec k}+ B_{\vec k}\Big)
\right]^{1/2} ~.
\label{freq}
\ee
In the continuum limit ${\vec k}\to 0$, 
one finds the approximate analytic expression 
\bea
\epsilon_{\vec k} &\approx&  \Omega {\cal S}
\Big[ \phantom{\times} 
\Big(
\Delta_1(\phi)-{{4\pi k}\over {ab}}
+O(k^2)
\Big) \nonumber \\
&&\phantom{\Omega {\cal S}\Big[} \times 
\Big(
\Delta_0(\phi) +
{{4\pi k}\over {ab}}
\cos^2 \theta_k
+O(k^2)
\Big)
\Big]^{1/2}
\label{cdc}
\eea
where $\Delta_0$ and $\Delta_1$, defined in Appendix C, depend on the
angle $\phi$. In particular, $\Delta_0$ vanishes for the highly symmetric
cases $\phi=\pi/3$ (triangular lattice) and $\phi=\pi/2$
(square lattice). 
The angle $\theta_k$ defines the orientation of the in-plane wavevector
${\vec k}$ with respect to the $\hat x$ axis.

It is worth stressing that $\epsilon_{\vec k}$ 
has the same structure both at zero and second order, 
because $\hdip^{(2)}$ does not contribute to the linear
term in $k$. In particular, we have that the linear term in
the quantity $(A_{\vec k}+B_{\vec k})$ vanishes when $\vec k$ is oriented
along the magnetization ($\theta_k=\pi/2$). 
Its sign is therefore decided by the quadratic term,
which is different for different lattices. It is positive for the triangular
lattice (see Fig.~4) and it may be either positive (Fig.~5b) or
negative (Fig.~5a) for the square lattice, depending on the orientation of
the magnetization.

\begin{figure}
\includegraphics[bbllx=170pt,bblly=70pt,bburx=470pt,bbury=700pt,
 width=7cm]{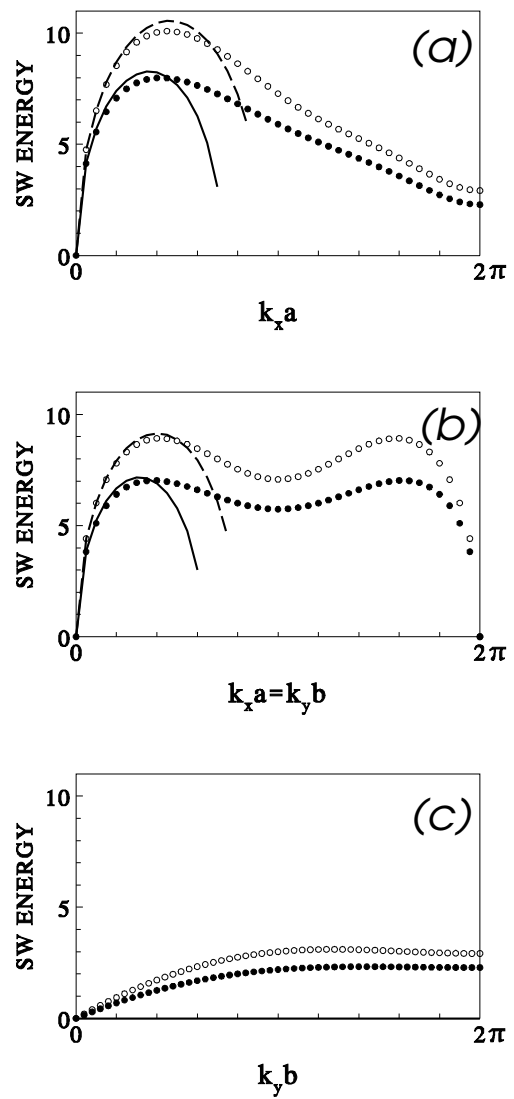}
\caption{Spin-wave dispersion curves 
calculated for the triangular lattice
(rhombic angle $\phi=\pi/3$) both numerically (Eq.~(\ref{freq}), symbols)
and in the continuum limit (Eq.~(\ref{cdc}), lines).
Solid circles and full lines refer to $\hdip^{(0)}$;
open circles and dashed lines refer to $(\hdip^{(0)}+\hdip^{(2)})$,
with ${\cal I}=0.1$ (units with $D_0=1$ are used).
The magnetization is assumed to lie along the $\hat y$ axis and
three different propagation directions are reported:
(a) $\theta_k=0$; (b) $\theta_k=\pi/6$; (c) $\theta_k=\pi/2$. 
Note the different periodicities of the spin-wave energy.
}
\end{figure}

\begin{figure}
\includegraphics[bbllx=135pt,bblly=80pt,bburx=550pt,bbury=690pt,
 width=7cm]{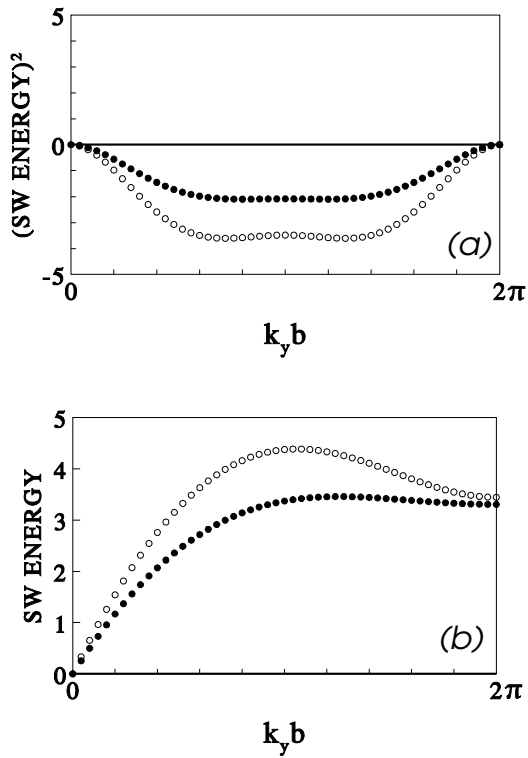} 
\caption{
The same as in Fig. 4, but for the square 
lattice (rhombic angle $\phi=\pi/2$).
(a) When magnetization and wavevector ${\vec k}$ are 
directed along a side of the square lattice, the square of 
the spin-wave energy is found to be negative, signaling 
the instability of the ferromagnetic configuration.
(b) When magnetization and wavevector ${\vec k}$ are directed along
the diagonal of the square, the ferromagnetic state is 
found to be metastable.
}
\end{figure}

In Figs.~4,5 we show the dispersion relation of the spin waves,
as obtained from Eq.~(\ref{freq}), in the case of a 
triangular lattice ($\phi=\pi/3$) and of a square one
($\phi=\pi/2$), for different orientations $\theta_k$.

In the remaining part of this Section we are making general comments on 
these results.

Dipolar interaction, since it couples the spins to the underlying
lattice, is not rotationally invariant, contrary to, 
{\it e.g.}, the Heisenberg
interaction (${\cal H}=-J\vec s_1\cdot\vec s_2$).
Therefore, the energy of a generic spin configuration depends on how
the spins are oriented with respect to the lattice. However, lattices with 
four-fold ($\phi=\pi/2$) or six-fold ($\phi=\pi/3$) symmetries are special,
in the sense that the energy of certain configurations 
(the ferromagnetic one, for example) is rotationally invariant.
This means that the dipolar ferromagnetic ground state of a
triangular lattice has a continuous degeneracy. 

The coupling between spins and crystal lattice manifests itself in two ways
(we are considering the highly symmetric cases, $\Delta_0(\phi)=0$).
First, the spin-wave energy $\epsilon_{\vec k}$ depends on
the orientation of the wavevector in the continuum limit
${\vec k}\to 0$ as well. According to Eq.~(\ref{cdc}),
$\epsilon_{\vec k}\sim |\cos\theta_k|k^{1/2}$, showing 
that the $\theta_k$-dependence of $\epsilon_{\vec k}$ is maintained in the
limit $\vec k\to 0$.

Second, even if the energy of the FM state is degenerate with
respect to its orientation in the plane, the spin-wave energy is 
{\it not} (see Fig.~5). 
This is true for the triangular lattice as well (not shown),
but it has most striking consequences for the square lattice.
In this case, the FM configuration is known to have a higher energy than 
the structure where spins are ferromagnetically coupled along lines and
antiferromagnetically coupled between neighbouring lines.\cite{Roz91}
The FM state, if oriented along an axis of the square lattice
(the typical configuration considered in the literature) is not even
locally stable. This is clearly shown in Fig.~5a where the square 
of the spin-wave energy is plotted, in the case of $\vec k$ and $\vec M$
parallel to the a side of the square lattice.
The square of the energy is negative,
signaling that the corresponding FM state is unstable.
However, if the magnetization is oriented along the diagonal, the
FM state is no more unstable, as shown in Fig.~5b (we show just the case
of $\vec k$ parallel to $\vec M$, but no instability appears for any
value of $\theta_k$).

Figures 4 and 5 display the effect of $\edip^{(2)}$  on the spin-wave spectrum:
the energy increases for any value and orientation of ${\vec k}$.
This result corroborates our statement that the finite size 
of the particles reinforces ferromagnetism in the triangular lattice
(Fig.~4). For the square lattice we have showed that ferromagnetism is
locally stable if $\vec M$ is oriented along a diagonal
of the square lattice (Fig.~5b) and locally unstable if $\vec M$ is parallel
to a side of the square lattice (Fig.~5a): 
$\edip^{(2)}$ reinforces the stability in the former case and the
instability in the latter one.

\section{The Curie temperature}
\label{sec:curie_temp}

The Curie transition temperature 
of an orthorombic lattice of particles 
with magnetization directed in plane along the ${\hat y}$ axis
can be estimated in the framework of spin-wave theory.
The relative deviation of the magnetization from the
saturation value takes the form\cite{Maleev} 
\bea
{{\delta {\cal S}}\over {\cal S}}&=&{1\over {N_{\Vert}{\cal S}}}
\sum_{\vec k} \langle 
a_{\vec k}^{\dagger} a_{\vec k}
\rangle= 
{{{\cal V}_2}\over {(2\pi)^2 {\cal S}}} 
\nonumber \\
&\times& 
\int d^2 {\vec k} 
\left[
{ A_{\vec k}\over \epsilon_{\vec k} }
{1\over {\rm e}^{ \epsilon_{\vec k}/T}-1 } -
{ A_{\vec k} - \epsilon_{\vec k} \over 2\epsilon_{\vec k} }
\right]
\label{deviation}
\eea
\noindent where ${\cal V}_2$ is the volume of the two-dimensional 
unit cell and the integration is over the first Brillouin zone.
The first term on the r.h.s. of Eq.~(\ref{deviation})
gives the temperature dependence of the magnetization, while
the second term represents the zero-point spin deviation,
which can be safely neglected.
A rough estimate of the Curie temperature $T_C$ 
is obtained by imposing that
$\delta {\cal S}(T_C)/{\cal S} \approx  1$.

Since the only energy scale in the problem is given by the dipolar
interaction, $T_C$ is expected to be of order 
$\Weff=\Omega {\cal S}^2/D_0^3
\approx (g\mu_B s)^2(L/a_0)^4 (t/c_0)^2/D_0^3$,
the effective dipolar interaction
between different particles.   

The convergence of the integral (\ref{deviation}) in
$\vec k=0$ is easily proved.\cite{Malo}
For generic orthorombic lattices, it is guaranteed by the gap
in the dispersion curve, $\epsilon_0\ne 0$.
For square and triangular lattices, 
the spin-wave energy vanishes for $k \to 0$ and
one can expand the exponential on the denominator 
of Eq.(\ref{deviation}), because the possible divergence is
infrared-like. Thus, the integral giving 
the temperature dependent spin deviation
is found to converge, provided that the {\it positive}~\cite{nota} 
$O(k^2)$ terms 
in the spin-wave energy are taken into account.
Eq.~(\ref{deviation}) can be rewritten as follows:
\be
{{\delta {\cal S}}\over {\cal S}}
\approx {{T}\over {\Weff}}
\int_0^{2\pi} d\theta
\int_0^{q_M}
{q dq\over {c_1 q \cos^2 \theta +c_2 q^2}}
\ee
\noindent where $q=k D_0$, $\theta=\theta_k$,
$c_1$ and $c_2$ are positive constants, and $q_M\approx 1$.
The double integral can be easily evaluated, giving
\be
T_C \approx \Weff \cdot {c_2\over 2\pi} 
\left[ \ln\left( {q_M + \sqrt{q_M^2 +(c_1/c_2)^2 } \over c_1/c_2 }
\right) \right]^{-1} \; .
\ee

In the limit $c_1=0$, $\epsilon_{\vec k}=c_2 q^2$ reproduces the
dispersion curve of the Heisenberg ferromagnet and the Curie
temperature $T_C$ vanishes, in agreement with the
Mermin-Wagner theorem.\cite{MW}

Finally, it is interesting to discuss the 
case where a uniaxial single-ion anisotropy 
$\lambda$, favouring the in-plane $\hat y$ axis, is present in the system. 
The effect of the finite size of the particle is straightforward,
in this case:
\be
\eani = -\lambda {\cal N} \sum_{{\vec D}_i} (s_i^y)^2 =
-\tilde\lambda \sum_{{\vec D}_i} ({\cal S}_i^y)^2
\ee
where $\tilde\lambda =\lambda/{\cal N}$ and, as usual, $\vec{\cal S} =
{\cal N}\vec s$.

The effect of $\eani$ on the spin-wave energy is simply 
to add the quantity $2\lambda$ to $A_{\vec k}$;
for large particles, this constant
factor dominates the dipolar terms in Eq.~(\ref{cdc}).
Because of that, spin-wave approximation is no more suitable and 
Eq.~(\ref{deviation}) can not be used to evaluate $T_C$:
in fact, it would give a Curie temperature $T_C\approx 2{\cal N}\lambda$,
which diverges when $\lambda\to\infty$.

On physical grounds, we expect that $T_C$ is an increasing function of 
$\lambda$, but in the limit of strong anisotropy, $T_C$ is always
of order $\Weff$. An analogy can be done with the three dimensional
Heisenberg model in the presence of an easy-axis anisotropy $\lambda$:
$T_C$ increases with $\lambda$, but $T_C(0)$ and $T_C(\infty)$ are of the
same order of magnitude and are both of order $J$, the exchange coupling
constant. In our case, $\Weff$ replaces $J$ and the dimension of the system
is two instead of three (the two-dimensional
Heisenberg model is not ordered at finite
temperature in the absence of anisotropy).

In order to corroborate our argument,
$T_C$ will be calculated in the mean field approximation, starting from 
the Hamiltonian 
\be
{\cal H}^{MF}=\sum_{{\vec D}_i} 
\left\lbrack~ \xi {\cal S}_i^y 
-\tilde\lambda 
\big( {\cal S}_i^y\big)^2
\right\rbrack
\ee
\noindent where $\xi={1\over 2}\Omega \langle
{\cal S}^y \rangle {\cal D}^{(0)}_{yy}(0)$.
The magnetization is given by 
\be
\langle {\cal S}^y \rangle=
{{
\sum_{M=-{\cal S}}^{{\cal S}} M 
{\rm e}^{-(\xi M-\tilde\lambda M^2)/T}
}\over {
\sum_{M=-{\cal S}}^{{\cal S}} 
{\rm e}^{-(\xi M-\tilde\lambda M^2)/T}
}}
\ee
\noindent where $M$, $M^2$ denote the eigenstates of ${\cal S}^y$, 
$({\cal S}^y)^2$ respectively.
As $T \to T_C$ one has $\langle {\cal S}^y \rangle \to 0$ 
so that $\xi \to 0$ and the exponential can be expanded. Thus
$ \langle {\cal S}^y \rangle \approx -{{\xi}\over {T_C}} R$
where $R$ is defined as 
\be
R={{
\sum_{M=-{\cal S}}^{{\cal S}} M^2 {\rm e}^{{{\tilde\lambda M^2}\over {T_C}}}
}\over
{
\sum_{M=-{\cal S}}^{{\cal S}} {\rm e}^{{{\tilde\lambda M^2}\over {T_C}}}
}}=
\langle ({\cal S}^y)^2 \rangle_{T_C}.
\ee

For $\tilde\lambda=0$ one has $R={1\over 3}{\cal S}({\cal S}+1)$, so that
$T_C(\tilde\lambda=0)={1\over 6} \Omega [-{\cal D}^{(0)}_{yy}(0)] 
{\cal S}({\cal S}+1)\approx \Weff$. 

For $\tilde\lambda \ne 0$ the mean field critical temperature is
\be
{{T_C(\tilde\lambda)}\over {T_C(\tilde\lambda=0)}} =
{3\over {{\cal S}({\cal S}+1)}}
\langle ({\cal S}^y)^2 \rangle_{T_C} ~,
\label{general}
\ee
\noindent where the mean on the r.h.s. must be calulated numerically.
Now we observe that in the limit $\tilde\lambda \to \infty$, one has 
$\langle ({\cal S}^y)^2 \rangle_{T_C} \to {\cal S}^2$, so that 
the ratio $T_C(\infty)/T_C(0)$ 
tends to the finite value $3{\cal S}/({\cal S}+1)$.
We conclude that for any value of $\tilde\lambda$, the Curie transition
temperature remains of the order of the effective dipolar 
interaction $\Weff$, with a prefactor changing by 
a factor three as $\tilde\lambda$ increases from 0 to $\infty$.

\section{Discussion}
\label{sec:concl}

Let us get back to the three questions formulated at the end of the 
Introduction.

{\it i) What is the effective dipolar interaction between single domain 
particles?} 

In the hypothesis that spins are strongly coupled ferromagnetically
inside each particle, it is straightforward to define an effective dipolar
coupling between (microscopic) spins $\vec s$ of 
two particles at distance $D$:
$\Weff=\Omega (s {\cal N})^2/D^3$, where $\Omega=(g\mu_B)^2$ and
${\cal N}$ is the number of spins in each particle. 
If $L$ and $t$ are respectively its linear size and 
thickness, $\Weff\approx (g\mu_B s)^2 (L/a_0)^4 (t/c_0)^2/D^3$.
If a single-ion anisotropy $\lambda$ is present, its effective
value is just~\cite{nota_lambda} 
$\leff=\lambda {\cal N}\approx \lambda (L/a_0)^2 (t/c_0)$. 

The full dipolar interaction between particles can be expanded in
(even) powers of $(L/D)^m$. The first correction term ($m=2$)
gives an interaction decaying as $1/D^5$ with the distance between
particles. A couple of remarks are in order here. 
First, the `purely' spin term
in the dipolar interaction ({\it i.e.}, the term not coupled to the lattice)
is no more rotationally invariant: the $z$-components are more strongly 
coupled than in-plane components. Second, $\edip^{(2)}$ preserves
two important features: it is minimized when spins are aligned 
ferromagnetically along the joining line, and the dipolar field generated
by a `continuum' line of spins aligned along the line in a point
outside the line, vanishes.

{\it ii) What is the dipolar ground state of an ordered array of 
magnetic particles?}

We have considered the class of orthorombic lattices, which comprises
the triangular and the square lattices.
It is known that in the case of a lattice of spins,
the ground state for the six-fold symmetry is
ferromagnetic and for the four-fold symmetry has zero net magnetization.
These results are not modified when single-domain particles replace
single spins and $\edip^{(2)}$ is considered in addition to
$\edip^{(0)}$. The effect of $\edip^{(2)}$ is to reinforce, in some
sense, the effect of $\edip^{(0)}$. In particular, therefore,
a triangular lattice of two dimensional particles interacting through
the dipolar interaction has a ferromagnetic ground state.

{\it iii) What is the finite temperature behaviour of a in-plane
ferromagnetically ordered array of particles?}

In two dimensional systems, long range order at finite temperature 
is not certain: however, the long range dipolar forces are known to grant it. 
In this respect, $\edip^{(2)}$ is of minor importance, 
because it decays as $1/D^5$ and consequently the dipolar sums in the 
$\vec k$-space do not contribute to the terms linear in $\vec k$,
but to the quadratic terms only.

In the absence of anisotropies, it is elementary that the Curie
temperature $T_C$ is of order of the effective dipolar coupling,
$T_C\approx \Weff$, because it is the only energy scale in the
problem. If in-plane easy-axis anisotropies are present, we have
shown that $T_C$ is expected to increase, but not to change in order
of magnitude: according to mean field theory, $T_C$ increases by a factor
three passing from the `weak' anisotropy regime into the `strong'
anisotropy regime.

Denoting by $w=(g \mu_B s)^2/a_0^3$ the dipolar coupling between 
microscopic spins on a two-dimensional lattice with atomic distance $a_0$,
the effective dipolar interaction between particles 
of linear dimension $L$ and thickness $t$ at distance $D_0$ can
be rewritten as $\Weff =w (L/a_0)^4 (t/c_0)^2/(D_0/a_0)^3$. Thus,
the Curie temperature $T_C \approx \Weff$ of an ensemble 
of magnetic particles may be significantly larger 
than the Curie temperature of 
a two-dimensional lattice of 
microscopic spins, which is of order $w$. 
However, $L$ cannot be made larger than $D_0$, because $D_0$ scales with $L$:
at the best, therefore, $T_C\approx w (L/a_0) (t/c_0)^2$.

Even if we have no definite evaluation of the numerical
prefactor appearing in the previous estimate for $T_C$, we
suggest that regular arrays of two-dimensional particles magnetized
in plane might sustain long range order at experimentally relevant
temperatures (see also Ref.~\onlinecite{JP}).
However, this collective phenomenon may be masked by the
superparamagnetic behaviour of the single particle, appearing
below the blocking temperature 
$T_B \approx \leff\approx \lambda (L/a_0)^2 (t/c_0)$.

The condition $T_B > T_C$ is equivalent to $\lambda/w >\eta L^2 t/D_0^3$,
where $\eta$ is an unknown numerical factor.
If such a condition is satisfied, the Curie phase transition
at $T=T_C$ is not visible because thermodinamic equilibrium cannot be
attained below $T_B$. So, if we decrease temperature from the high-$T$
region each particle becomes superparamagnetic when dipolar forces
are still unable to induce a long range order in the system.
When $T=T_C$ dipolar interaction comes into play, but the magnetization
of each particle is frozen  and the system is unable to attain equilibrium. 
In the opposite case, $T_C > T_B$, a phase transition at $T_C$ should
be visible.

Recently, the system Co/Cu(001) has drawn the attention because
it has been suggested~\cite{Co_JMMM,Co_PRB} that for
$t<1.8$ML (ML=mo\-no\-layer) this system displays a dipolar
induced ferromagnetic order. $T_C(t)$ is seen~\cite{Co_JMMM} 
to be finite and increase
from $T_C$(1ML)=25K to $T_C$(1.8ML)=200K, where it has a sudden jump,
attributed to the percolation in the {\it second} layer.

The growth morphology of the system Co/Cu(001) is complicated by the 
alloying~\cite{Co_SS,Co_PRB} between the two elements, which mainly
takes place in the first layer (75$\%$ in Co, 25$\%$ in Cu).
However, with such a high percentage of Cobalt,
it is hard to suppose that the first layer is made up of an ensemble
of disconnected Co-islands which interact only through long range forces
(in Ref.~\onlinecite{Co_PRB} the percolation threshold is 
theoretically estimated to be of order 60$\%$). 
More likely, an infinite cluster of Cobalt does exist in the
first layer and therefore the magnetic behaviour of the system (and the
value of $T_C$) follows from the combining effect~\cite{Co_PRB} 
of the direct exchange interaction between spins and long range forces.

Finally, we would like to mention an additional difficulty in the
interpretation of experimental data concerning an array of magnetic 
islands: the random character of deposition and diffusion 
gives rise to a non-uniform distribution of sizes and positions.\cite{JP}
This fact, along with frustration due to dipolar interaction,
makes difficult even the determination of the ground state,
because the system has a glassy behaviour with a lot of metastable states.

\appendix

\section{Multipolar expansion of interparticle interaction}
\label{app1}

We start from Eq.~(\ref{edip}), assuming single monolayers.\cite{nota_spessore} 
Summations on $\vec R_{1,2}$
are replaced by sums on $\vec r_{1,2}$ (see Fig.~1 for notations):
\be
\edip = {1\over 2}\Omega \sum_{\vec r_1} \sum_{\vec r_2}  \left[
{ \su\cdot\sd\over R_{12}^3 }
-3 { (\su\cdot\vec R_{12})(\sd\cdot\vec R_{12})\over R_{12}^5 } \right]
\nonumber
\ee   

We make the expansions:
\be
{1\over R_{12}^3} \approx {1\over D^3} \left[
1 -3 {\vec r\cdot\vec D \over D^2} - {3\over 2}{r^2\over D^2}
+ {15\over 2}{(\vec r\cdot\vec D)^2\over D^4} \right]
\ee

\begin{widetext}
\bea
{ (\su\cdot\vec R_{12})(\sd\cdot\vec R_{12}) \over R_{12}^5 } \approx
&& {1\over D^5} \left\{ 
 (\su\cdot\vec D)(\sd\cdot\vec D)
\left[ 1 -5 {\vec r\cdot\vec D \over D^2} - {5\over 2}{r^2\over D^2}
+ {35\over 2}{(\vec r\cdot\vec D)^2\over D^4} \right]
+ (\su\cdot\vec r)(\sd\cdot\vec r)\right. \nonumber \\
&+& \left. [ (\su\cdot\vec D)(\sd\cdot\vec r) + 
(\sd\cdot\vec D)(\su\cdot\vec r)]
\left[ 1 -5 {\vec r\cdot\vec D \over D^2}\right] \right\}
\eea
\end{widetext}

The following expressions are easily calculated:
\bea
&& \sum_{\vec r_1} 1 = {\cal N}_1  ~~~ \sum_{\vec r_2} 1 = {\cal N}_2 
\nonumber \\
&& \sum_{\vec r_1} \sum_{\vec r_2} \vec r = {\cal N}_1 {\cal N}_2 
(\langle\vec r_2\rangle - \langle\vec r_1\rangle) \nonumber \\
&&  \sum_{\vec r_1} \sum_{\vec r_2} r^2 = {\cal N}_1 {\cal N}_2
(\langle r_1^2\rangle + \langle r_2^2\rangle -2 \langle\vec r_1\rangle
\cdot \langle\vec r_2\rangle) \nonumber \\
&& \sum_{\vec r_1} \sum_{\vec r_2}  
(\vec V_1\cdot\vec r)(\vec V_2\cdot\vec r) 
 = {\cal N}_1 {\cal N}_2  [ 
\langle(\vec V_1\cdot\vec r_1)(\vec V_2\cdot\vec r_1)\rangle \nonumber \\
&& + \langle(\vec V_1\cdot\vec r_2)(\vec V_2\cdot\vec r_2)\rangle
- (\vec V_1\cdot \langle\vec r_2\rangle)(\vec V_2\cdot \langle\vec r_1\rangle)
\nonumber \\
&& - (\vec V_1\cdot \langle\vec r_1\rangle)(\vec V_2\cdot 
\langle\vec r_2\rangle) ]  \nonumber 
\eea

In the previous expressions, ${\cal N}_{1,2}$ are the number of spins in 
the particles $I_{1,2}$ and $\vec V_{1,2}$ are generic vectors.

It is always possible to choose the origin of the reference system for
a given particle in its center of mass, so that $\langle \vec r_i\rangle=0$
and the ``moment of inertia" ${\cal I}_i$ of a particle is
${\cal I}_i = \langle r_i^2 \rangle $.
Consequently, we have the following results
\bea
\langle (\vec D \cdot \vec r_i)^2 \rangle &=&  
\fra{1}{2} D^2 {\cal I}_i \\
\langle (\vec V_1 \cdot \vec r_i)(\vec V_2 \cdot \vec r_i) \rangle &=&
\fra{1}{2} {\cal I}_i \vec V_1^\parallel \cdot \vec V_2^\parallel
\eea
where $\vec V^\parallel$ is the in-plane component of the
generic vector $\vec V$.

If we define ${\cal I}_{12}=\fra{1}{2}({\cal I}_1+{\cal I}_2)$ and
$ \vec {\cal S}_i = {\cal N}_i \vec s_i$,  we obtain 
the following expression for the effective dipolar interaction between two
particles at distance $\vec D$:
\bea
\edip = && {1\over 2}\Omega { \Su^\perp \cdot \Sd^\perp \over D^3}
\left( 1 + {9\over 2}{ {\cal I}_{12}\over D^2 } \right) \\
 &+& {1\over 2}\Omega { \Su^\parallel \cdot \Sd^\parallel \over D^3}
\left( 1 + {3\over 2}{ {\cal I}_{12}\over D^2 } \right) \\ 
 &-& {3\over 2}\Omega 
{ (\Su^\parallel\cdot\vec D)(\Sd^\parallel\cdot\vec D)\over
D^5 } \left( 1 + {5\over 2}{ {\cal I}_{12}\over D^2 } \right) 
\eea

\section{Minimization of the dipolar coupling between two spins}
\label{app2}

Let us consider two unitary spins $\vec S_1,\vec S_2$ located 
at a distance $R_{12}$ along the in-plane $\hat x$ axis,
taken as the polar axis, while the $\hat z$ axis is perpendicular to the 
plane. Their orientations are defined by the
polar and azimuthal angles $\theta_i,\varphi_i$.

The dipolar coupling can be generally written as
\bea
E_{12} &\approx& \phantom{+}E_{12}^{(0)} + E_{12}^{(2)} \nonumber\\
&\approx& \phantom{+}\tilde\Omega_0 (\vec S_1\cdot\vec S_2 -3 S_1^x S_2^x
)\nonumber\\
&& +\tilde\Omega_2 (2S_1^z S_2^z + \vec S_1\cdot\vec S_2 -5 S_1^x S_2^x )
\nonumber  
\eea
where $\tilde \Omega_0 ={1\over 2} \Omega S^2/R_{12}^3$ and
$\tilde\Omega_2 = \fra{3}{4}{\cal I}_{12}\Omega S^2/R_{12}^5$
(see Eqs.~\ref{e0},\ref{e2}).

Both terms $E_{12}^{(0)}$ and $E_{12}^{(2)}$ are minimized by
the same configuration. First, let us treat the zero-order term.
We have to minimize the function
\be
E_{12}^{(0)}/\tilde\Omega_0 = \sin\theta_1\sin\theta_2\cos(\varphi_1-\varphi_2)
- 2\cos\theta_1\cos\theta_2
\ee

By taking the derivatives with respect to $\varphi_{1,2}$ we
find that $(\varphi_1-\varphi_2)=0,\pi$ or that one $\theta_i$ at least
must vanish. If, {\it e.g.}, $\theta_1=0$, it is straightforward to
derive that $\theta_2=0$ as well. If both $\theta_{1,2}$ are not
vanishing, taking the derivatives with respect to them
implies $\cos\theta_1=\cos\theta_2=0$, {\it i.e.} $\theta_{1,2}=\fra{\pi}{2}$.
In simple words, it is sufficient to consider two kinds of configurations:
i)~the configuration where both spins are perpendicular to the joining
vector and they are parallel or antiparallel;
ii)~the configuration where both spins are aligned along the joining
vector. The former configuration, with antiparallel spins, corresponds
to the minimization of $(\vec S_1\cdot\vec S_2)$ and its energy is
$-\tilde \Omega_0$. 
The latter configuration, with parallel spins, corresponds
to the minimization of $(-3 S_1^x S_2^x)$ and its energy is
$\tilde\Omega_0 -3\tilde\Omega_0= -2\tilde\Omega_0$.
We conclude that $E_{12}^{(0)}$ is minimized by the ferromagnetic 
configuration with spins aligned along their joining vector.

The minimization of $E_{12}^{(2)}$ proceeds along the same lines,
with the minor difference that configurations with both spins perpendicular
to the joining line are no more degenerate with respect to a global
rotation around the $\hat x$ axis: the lowest energy one corresponds
to antiparallel spins along the $\hat z$ direction and its energy is
$-3\tilde\Omega_2$. On the other hand, the ferromagnetic configuration
with both spins parallel to the $\hat x$ axis has the energy 
$-4\tilde\Omega_2$, so the conclusion is unchanged.

\section{Dipolar sums for the orthorombic lattice}
\label{app3}

For spins ferromagnetically oriented along the $y$ axis, 
the coefficients of the spin-wave Hamiltonian 
in Eq.~(\ref{spinwave}) take the form
\bea
{A^{(0)}_{\vec k}\over {\Omega {\cal S}}}&=&
\left[ 
{1\over 2} {\cal D}^{(0)}_{zz}({\vec k}) 
+{1\over 2}{\cal D}^{(0)}_{xx}({\vec k})
-{\cal D}^{(0)}_{yy}(0)
\right]
\nonumber \\
{B^{(0)}_{\vec k}\over {\Omega {\cal S}}}&=&
\left[
{1\over 2} {\cal D}^{(0)}_{zz}({\vec k}) 
- {1\over 2}{\cal D}^{(0)}_{xx}({\vec k})
+i {\cal D}^{(0)}_{zx}({\vec k})
\right]
\label{sp0}
\eea
and
\bea
{A^{(2)}_{\vec k}\over
{\Omega  {\cal S} {3\over 2} {\cal I}}}&=&
\left[ 
{1\over 2} {\cal D}^{(2)}_{zz}({\vec k}) 
+{1\over 2}{\cal D}^{(2)}_{xx}({\vec k})
-{\cal D}^{(2)}_{yy}(0)
+{\cal E}^{(2)}({\vec k})
\right]
\nonumber \\
{B^{(2)}_{\vec k}\over 
{\Omega  {\cal S} {3\over 2} {\cal I}}}&=&
\left[
{1\over 2} {\cal D}^{(2)}_{zz}({\vec k}) 
- {1\over 2}{\cal D}^{(2)}_{xx}({\vec k})
+i{\cal D}^{(2)}_{zx}({\vec k})
+{\cal E}^{(2)}({\vec k})
\right]
\nonumber \\
\label{sp2}
\eea
\noindent where ${\cal N}$ is the number of spins 
in each particle and the dipolar sums 
are defined as ($\alpha,\beta=x,y,z$)
\bea
{\cal D}^{(n)}_{\alpha \beta}({\vec k})&=&\sum_{{\vec D}_j} 
{1\over {D_{ij}^{3+n}}} \left[ 1- (3+n) 
{ {D_{ij}^{\alpha}D_{ij}^{\beta} }\over 
{D_{ij}^2}}
\right] 
{\rm e}^{i {\vec k} \cdot {\vec D}_{ij}}
\nonumber \\
{\cal E}^{(n)}({\vec k})&=&\sum_{{\vec D}_j} 
{{
{\rm e}^{i {\vec k}\cdot {\vec D}_{ij}} }
\over {D_{ij}^{3+n}} }
\eea
We observe that in the ultraflat particle limit, 
one has 
${\cal E}^{(n)}({\vec k}) \approx D_{zz}^{(n)}({\vec k})$ 
and $D_{zx}^{(n)}({\vec k})\approx 0$ and the
${\cal D}_{\alpha\alpha}$'s can be approximately expressed 
\bea
{\cal D}^{(n)}_{xx}({\vec k}) &\approx&
Y^{(n)}({\vec k})-(2+n)X^{(n)}({\vec k})
\nonumber \\
{\cal D}^{(n)}_{yy}({\vec k}) &\approx&
X{(n)}({\vec k})-(2+n)Y^{(n)}({\vec k})
\nonumber \\
{\cal D}^{(n)}_{zz}({\vec k}) &\approx&
X^{(n)}({\vec k})+Y^{(n)}({\vec k})
\eea
in terms of the dipolar sums
\bea
{X}^{(n)}({\vec k})&=&\sum_{{\vec D}_{j}} 
 {{(D_{ij}^{x})^2}\over D_{ij}^{5+n}}
~{\rm e}^{i {\vec k} \cdot {\vec D}_{ij}}
\nonumber \\
{Y}^{(n)}({\vec k})&=&\sum_{{\vec D}_j} 
{{(D_{ij}^{y})^2}\over D_{ij}^{5+n}}
~{\rm e}^{i {\vec k} \cdot {\vec D}_{ij}}.
\eea
The latter sums can be numerically calculated in a very efficient way 
following a method, developed some years ago by Benson and Mills,\cite{BM}
similar to Ewald's one for the evaluation of lattice sums. 
For $n=0$ one has
\be
{X}^{(0)}(k_x,k_y;a,b)=
{U}^{(0)}(k_x,k_y)+
{V}^{(0)}(k_x,k_y)
\ee
where ${U}$ is a summation over the sites of the 
orthorombic lattice which are located at integer
multiples of $a$ and $b$ 
\bea
{U}^{(0)}(k_x,k_y)&=&
{{16}\over {3}}{1\over {b}}
\sum_{l=1}^{+\infty} \cos(k_x a l)
\nonumber \\ 
\sum_{m=-\infty}^{+\infty}({{\pi m}\over {b}}&+&{{k_y}\over 2})^2
K_2\left[ 2 a l \vert {{\pi m}\over {b}}+{{k_y}\over 2}\vert
\right]
\nonumber
\eea
while ${V}$ refers to semi-integer multiples
\bea
{V}^{(0)}(k_x,k_y)&=&
{{16}\over {3}}{1\over {b}}
\sum_{l=0}^{+\infty} \cos\left[ k_x a (l+{1\over 2})\right]
\nonumber \\ 
\sum_{m=-\infty}^{+\infty}(-1)^m
({{\pi m}\over {b}}&+&{{k_y}\over 2})^2
K_2\left[ 2 a (l+{1\over 2}) 
\vert {{\pi m}\over {b}}+{{k_y}\over 2}\vert
\right]
\nonumber
\eea
For $n=2$ one has
\be
{X}^{(2)}(k_x,k_y;a,b)=
{U}^{(2)}(k_x,k_y)+
{V}^{(2)}(k_x,k_y)
\ee
where
\bea
{U}^{(2)}(k_x,k_y)&=&
{{32}\over {15}}{1\over {ab}}
\sum_{l=1}^{+\infty} {{\cos(k_x a l)}\over {l}}
\nonumber \\ 
\sum_{m=-\infty}^{+\infty}({{\pi m}\over {b}}&+&{{k_y}\over 2})^3
K_3\left[ 2 a l \vert {{\pi m}\over {b}}+{{k_y}\over 2}\vert
\right]
\nonumber
\eea
and
\bea
{V}^{(2)}(k_x,k_y)&=&
{{32}\over {15}}{1\over {ab}}
\sum_{l=0}^{+\infty} {{\cos\left[ k_x a (l+{1\over 2})\right]}\over 
{(l+{1\over 2})}}
\nonumber \\ 
\sum_{m=-\infty}^{+\infty}(-1)^m
({{\pi m}\over {b}}&+&{{k_y}\over 2})^3
K_3\left[ 2a (l+{1\over 2}) 
\vert {{\pi m}\over {b}}+{{k_y}\over 2}\vert
\right]
\nonumber
\eea
Here above, $K_2(z)$ and $K_3(z)$ are the modified Bessel functions
of second and third order, respectively. By symmetry reasons, one has
\be
{Y}^{(n)}(k_x,k_y;a,b)=
{X}^{(n)}(k_y,k_x;b,a)
\ee
Finally, in the continuum limit
${\vec k}\to 0$, one can obtain
analytical expressions for the dipolar sums. 
Denoting by $k$ the modulus of the two-dimensional
wavevector ${\vec k}=(k_x,k_y)$
and by $\theta_k$ the angle that ${\vec k}$ forms with
the ${\hat x}$ axis, we obtain 
\bea
{X}^{(0)}(k)&\approx&{X}^{(0)}(0)-
{{2\pi}\over {ab}}~ k \left[
1+ {1\over 3} \cos(2\theta_k)\right]
\nonumber \\
{Y}^{(0)}(k)&\approx&{Y}^{(0)}(0)-
{{2\pi}\over {ab}}~ k \left[
1- {1\over 3} \cos(2\theta_k)\right]
\eea
and
\bea
X^{(2)}(k) &\approx& X^{(2)}(0)+O(k^2)
\nonumber \\
Y^{(2)}(k) &\approx& Y^{(2)}(0)+O(k^2)
\eea
where for the triangular lattice one has
$X^{(0)}(0)=5.5170879/D_0^3$, $X^{(2)}(0)=3.3809493/D_0^3$
and for the square lattice $X^{(0)}(0)=4.5168109/D_0^3$, 
$X^{(2)}(0)=2.5451291/D_0^3$.
From the previous expressions one obtains
approximate expansions for the quantities 
in Eqs.~(\ref{freq},\ref{deviation}):
\bea
{{A_{\vec k}-B_{\vec k}}\over  {\Omega {\cal S}} } &\approx&
\Delta_{0}+{{4\pi k}\over {ab}} \cos^2 \theta_k
+O(k^2)
\nonumber \\
{{A_{\vec k}+B_{\vec k}}\over  {\Omega {\cal S}} } &\approx&
\Delta_1-{{4\pi k}\over {ab}}
+O(k^2)
\nonumber \\
{A_{\vec k}\over 
{\Omega {\cal S}}}
&\approx& \Delta_2 
-{{2\pi k }\over {ab}} 
\sin^2 \theta_k +O(k^2)
\eea
where 
\bea
\Delta_0&=& 3\left[ Y^{(0)}(0)-X^{(0)}(0)\right]
+{{15}\over {2}} {\cal I}
\left[Y^{(2)}(0)-X^{(2)}(0) \right]
\nonumber \\
\Delta_1 &=& 3Y^{(0)}(0)+
{3\over 2}{\cal I}\left[7 Y^{(2)}(0)+2X^{(2)}(0)\right]
\nonumber \\
\Delta_2&=&{3\over 2}\left[ 2Y^{(0)}(0)-X^{(0)}(0)\right]
+ {9\over 4}{\cal I} \left[4 Y^{(2)}(0)-X^{(2)}(0)\right]
\nonumber \\
\eea
Hence we observe that for a generic orthorombic lattice 
one has $X^{(n)}(0) \ne Y^{(n)}(0)$, so that
the dispersion curve 
$\epsilon_{\vec k}=
\left[(A_{\vec k}-B_{\vec k})(A_{\vec k}+B_{\vec k})
\right]^{1/2}$
presents a gap for ${\vec k} \to 0$.
For the special cases of 
the triangular and the square lattice 
one has $X^{(n)}(0)=Y^{(n)}(0)$ by symmetry reasons;
this implies that $\Delta_0=0$, so that
a Goldstone mode is present
in the dispersion curve: $\epsilon_{\vec k} \to 0$ for
${\vec k} \to 0$.

\begin{acknowledgments}
We thank Angelo Rettori for his critical reading of the manuscript.
\end{acknowledgments}


\begin{thebibliography}{}
\bibitem{nota_lambda}
$\leff$ should not be mistaken for $\tilde\lambda$.
The former is the effective anisotropy  for a microscopic
spin $\vec s$, the latter for the giant spin $\vec {\cal S}$.
\bibitem{nota_spessore}
Let us remind that $\edip$ is just linear in $t_1/c_0$ and $t_2/c_0$.
\end{thebibliography}
\end{document}